\documentclass[
reprint,
amsmath,amssymb,
aps, %aps
]{revtex4-2}
\usepackage[utf8]{inputenc}
\usepackage{newunicodechar}
\newunicodechar{̈}{\"{}}

\usepackage{hyperref}% add hypertext capabilities
\hypersetup{
	colorlinks=true,       % false: boxed links; true: colored links
	linkcolor=blue,          % color of internal links
	citecolor=blue,        % color of links to bibliography
}
\usepackage{graphicx}% Include figure files
\usepackage{dcolumn}% Align table columns on decimal point
\usepackage{bm}% bold math
\usepackage{ragged2e} % for \justifying
\usepackage{array} % for advanced table column definitions

\usepackage{comment}
\usepackage{lipsum}
\usepackage{multirow}
\usepackage{booktabs} 
\renewcommand{\arraystretch}{1.5}
\usepackage{subcaption}
\usepackage{float}
\usepackage{soul}
\usepackage[font=small,labelfont=bf]{caption}
\usepackage{makecell}
\usepackage{array} % for p{} columns
\usepackage{geometry}
\usepackage{xr}
\usepackage{amsmath,graphicx}
\usepackage{adjustbox}   % for scalable box
\usepackage{makecell}    % for multi-line cells

\begin{document}

\preprint{APS/123-QED}
%\title{Power Laws Challenge Conventional Insights on Transient Photoluminescence in Perovskites}
\title{Power Law Transients in Perovskite Photoluminescence}
%\title{Tracing Power-Law Dynamics in Perovskite Transient Photoluminescence}

\author{Anas M. M.}
\email{anasmm99@gmail.com}
\affiliation{Department of Electrical Engineering, Indian Institute of Technology Bombay, Mumbai, India}

\author{Pradeep R. Nair}
\email{prnair@ee.iitb.ac.in}
\affiliation{Department of Electrical Engineering, Indian Institute of Technology Bombay, Mumbai, India}

\date{\today}% It is always \today, today,
\begin{abstract}

Carrier dynamics and recombination in perovskite thin films are often explored through transient photoluminescence (PL) measurements and subsequent analysis based on single/multi-exponentials. In contrast, here we show that several experimental PL transients reported in literature  vary as power laws with time and are not multi-exponentials. Puzzlingly, the existing theoretical methods are ill-equipped to explain the origin and implications of such power law trends. Our analysis resolves this puzzle and correlates the power law exponents with the trap distributions. Curiously, we show that PL transients could vary as $t^{-2}$ even in the presence of dominant monomolecular recombination while the existing theory associates this unique trend exclusively with dominant bimolecular recombination. These findings identify the limits of applicability of concepts such as differential lifetime and highlight the need for predictive modeling synergistic with insightful data analysis towards performance optimization of perovskite optoelectronic devices.

\end{abstract}
\maketitle

\section{\label{sec:intro}Introduction}
The performance of optoelectronic devices such as solar cells, photodetectors, and light emitting diodes is often limited by recombination processes \cite{goetz2020shining,hu2024flattening,yang2022defect,martani2023defect,ory2024effect,zhou202324,mahon2020photoluminescence,wei2019ultrafast,li2019direct}. Among several techniques, transient photoluminescence (TPL) is widely used for probing carrier dynamics and recombination pathways at material/thin film level \cite{chen2023characterizing,li2024defect,ulatowski2020charge,chen2024charge,stranks2014recombination,nair_2023,dreessen2024effective,huang2025decoding,stranks2026journey}. TPL, measured from photoactive thin films during pulsed illumination, is typically analyzed in terms of rate equations to explore the carrier dynamics. Several insights are readily available in this regard: For example, it is well known that PL transients vary exponentially with time if monomolecular process is the dominant recombination mechanism \cite{stranks2014recombination,bojtor2022ultralong}. On the other hand, the TPL could vary as $t^{-2}$ under dominant bimolecular recombination \cite{nandal2021unveiling,higashi2022design,tachiya2010theory}. In addition, multi-exponential fits are also regularly used to interpret TPL experiments\cite{elkhouly2025transient,chirvony2017delayed,chen2023study,thomaz2020single,chen2021carrier,peters2022photoluminescence,cao2022interband,kim2024unveiling,taddei2024interpreting,li2023diffusion,zhou2024role,dey2018role}. 

\begin{figure}[ht]
    \centering
    \captionsetup{labelformat=empty}
    \includegraphics[width=\columnwidth]{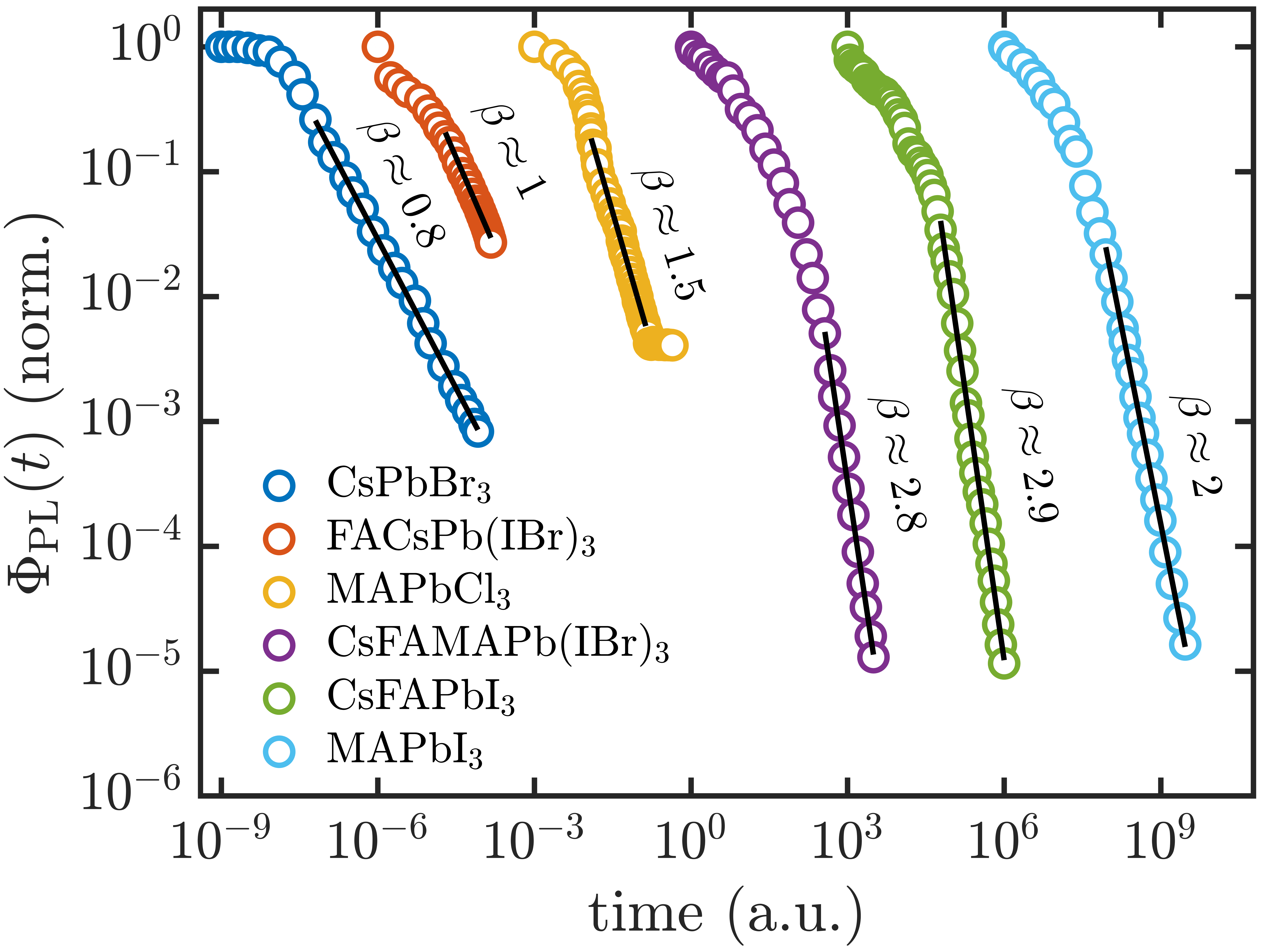}
    \captionsetup{font=small}

    \caption{\justifying
    \textbf{Figure~\ref{fig:experiment_combined}.}
     Experimental PL transients with $\beta \neq 2$ reported in the literature for different perovskite active layers. Each data set is time shifted (on a log-scale) for an uncluttered and better visualization of the power law nature of transients (see Sec. S1, Supporting Information, SI for plots in real time scale). The perovskite active materials and the corresponding references are:  $\mathrm{MAPbCl_{3}}$ \cite{becker2020unraveling}, $\mathrm{CsPbBr_{3}}$ \cite{oliver2022understanding}, $\mathrm{FACs(IBr)_{3}} $ \cite{li2024defect}, $\mathrm{CsFAPbI_{3}}$ \cite{yuan2025understanding}, $\mathrm{MAPbI_{3}}$ \cite{yuan2025understanding}, and $\mathrm{CsFAMAPb(IBr)_{3}}$ \cite{yuan2025understanding}.}

    \label{fig:experiment_combined}
\end{figure}

Indeed, the methodology to interpret TPL experiments is widely reported in the literature. As mentioned above, the transients are expected to vary as $e^{-t/\tau}$ or $t^{-2}$, where $\tau$ is a characteristic time constant. To check the consistency of this notion, we reviewed recent literature on perovskite photoluminescence (i.e., TPL measurements). As expected, we found that several experimental results show exponential variation with time. However, as shown in Figure \ref{fig:experiment_combined}, we find that, in many instances, the experimental PL transients vary as a power law with time (i.e.  $\Phi_{PL} (t) \propto t^{-\beta}$, where $\beta$ is the power law exponent). The power law trends extend over several orders of magnitude of PL transients. Interestingly, we find the power law exponent $\beta$ to be significantly different from $2$ in these literature reports.

The observation of power-law transients with $\beta \neq 2$ is a new territory as far as analysis of TPL is concerned. First, in spite of some recent reports \cite{yuan2025understanding}, it is not widely known that such non-integer power law exponents are possible in PL transients. As a result, it might be possible that such data are also wrongly analyzed and interpreted in terms of multi-exponentials. Second, the physical origin of such transients is unknown, Third, the implications of the same for perovskite optoelectronics is yet to be elucidiated. Fourth, it is not very evident how to interpret metrics like differential lifetime or quasi-Fermi level splits in view of such non-integer power law transients. All these indicate the necessity for a critical analysis of carrier dynamics and PL transients in perovskites.

In this context, here we show that non-integer power law transients are possible in perovskites due to the presence of trapping centers. The nature of trap distribution uniquely determines the power law exponents. Curiously, we also find that PL transients which vary as $t^{-2}$ are possible even in some select scenarios in which the monomolecular recombination dominates the bimolecular recombination - a possibility yet to be elucidated in literature. Further, our analyses also identify the contours of applicability of concepts like differential lifetime for perovskite optoelectronics. Below, we describe the model system followed by results and discussions.
\section{\label{sec:Three}Theory}
\begin{figure}[ht]
    \centering
    \captionsetup{labelformat=empty}

    \includegraphics[width=\columnwidth]{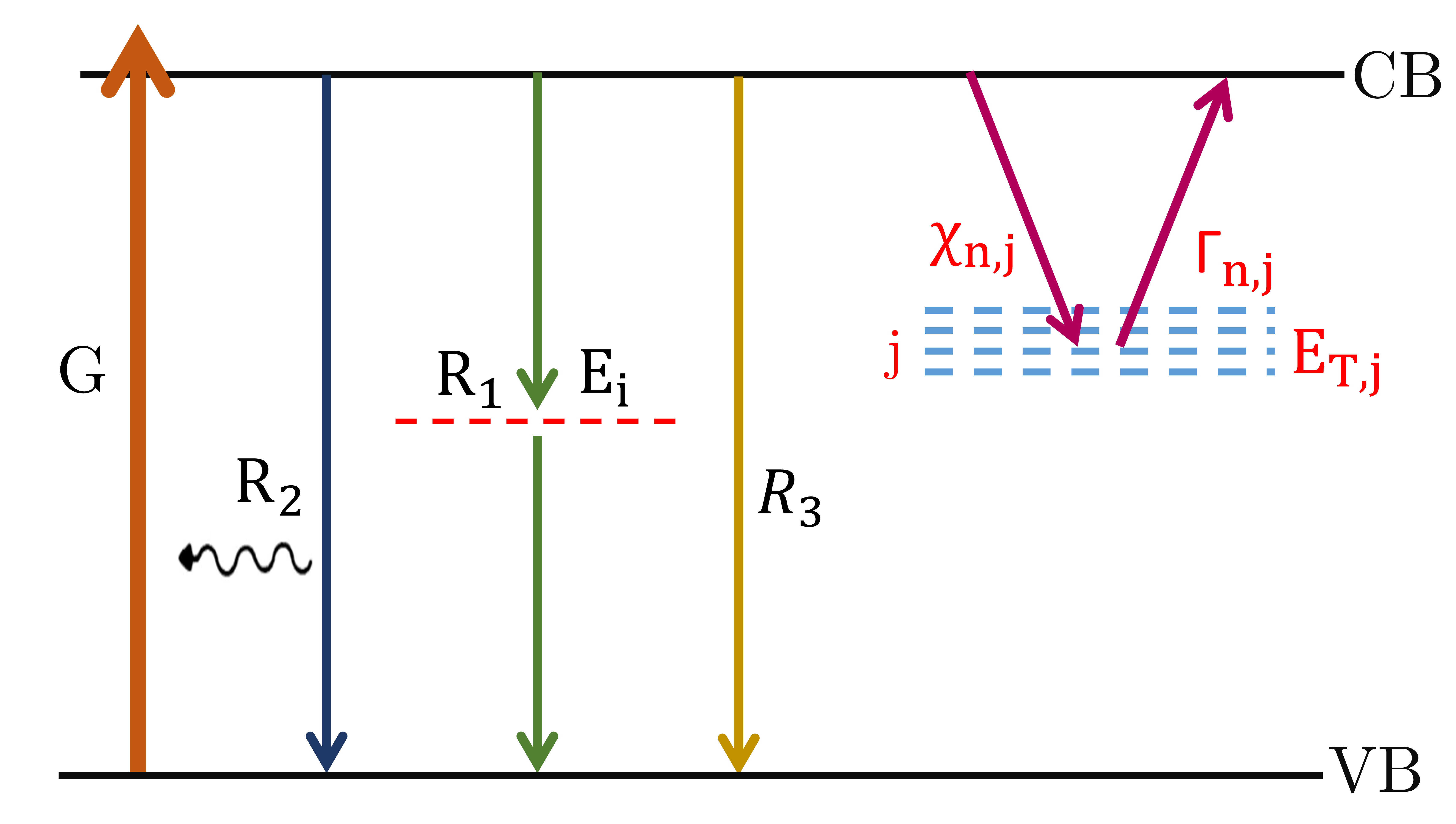}
 \captionsetup{font=small}
\caption{ \justifying { \textbf{Figure~\ref{Methodology}}.
    Schematic representation of carrier recombination and trapping processes. }}
    \label{Methodology}
\end{figure}

In TPL measurements, following short-pulse laser excitation, the photo-generated carrier population undergoes a complex interplay between radiative recombination, trap-assisted capture/emission, and non-radiative recombination. Figure \ref{Methodology} summarizes the various processes in such a measurement. Here $G$ denotes the carrier generation due to illumination, $R_1$ denotes the carrier recombination through mid-gap states, $R_2$ denotes the radiative recombination, and $R_3$ denotes the Auger recombination. Also indicated is a distribution of trapping centers which can trap and emit carriers. Such trapping centers do not contribute to carrier recombination, but could lead to distinct features in PL transients.\\

The temporal modulation of carrier populations under pulsed photo-excitation can be described using the system of coupled rate equations (eqs. \ref{eq1}-\ref{eq_p}). Here, \(n\) and \(p\) represent the concentrations of free electrons and holes, respectively, while \(n_{T}\) denotes the trapped electron density. The coefficients \(k_{1}\), \(k_{2}\), and \(k_{3}\) are the rate constants associated with monomolecular (trap assisted Shockeley-Read-Hall, SRH), bimolecular (radiative), and Auger (non-radiative) recombination, respectively\cite{stranks2014recombination}.

Figure \ref{Methodology} also depicts the carrier dynamics through a distribution of trapping centers\cite{nair_2023}. The subscript $j$ denotes trapping/emission at an energy level $E_{T,j}$ with $N_{T,j}$ being the density of traps. Eq. \ref{eq3} represents the rate of change of trapped electron density at energy level $E_{T,j}$. The electron capture rate at the same trap level is given by $\chi_{n,j}=c_n(N_{T,j}-n_{T,j})n$ while the electron emission rate to the conduction band is given as $\Gamma_{n,j}=c_nn_{T,j}n_{1,j}$ where $n_{1,j}=n_{i}e^{(E_{T,j}-E_i)/k_BT}$. Here \( E_i \) is the intrinsic Fermi level, \(n_i\) is the intrinsic carrier concentration, \( k_B \) is the Boltzmann constant, and \( T \) is the absolute temperature.

\begin{align}
\frac{\partial n}{\partial t} &= 
  G 
  - k_{1}n 
  - k_{2}(np - n_{i}^{2}) \nonumber\\
  &- k_{3}(n + p)(np - n_{i}^{2}) 
  - \sum_{j}(\chi_{n,j}-\Gamma_{n,j}) 
%  + \sum_{i=\text{min}}^{\text{max}} E_{n_{T}}(E_{T,i})
  \label{eq1} \\[6pt]
\frac{\partial p}{\partial t} &= 
  G 
  - k_{1}n 
  - k_{2}(np - n_{i}^{2})
  \nonumber\\
  &\quad 
  - k_{3}(n + p)(np - n_{i}^{2})
  \label{eq2} \\[6pt]
\frac{\partial n_{T,j}}{\partial t} &= 
\chi_{n,j}-\Gamma_{n,j}
 % \sum_{i=\text{min}}^{\text{max}} C_{n_{T}}(E_{T,i}) 
 % - \sum_{i=\text{min}}^{\text{max}} E_{n_{T}}(E_{T,i})
  \label{eq3}\\
  n_T&=\sum_{j}n_{T,j}
    \label{eq_nT}\\
 p&=n+n_T 
 \label{eq_p}
\end{align}

The observed photoluminescence intensity ($\Phi_{PL}$)  is governed by the radiative component, given by \(\Phi_{PL} = k_{2}(np - n_{i}^{2})\). Here, we provide analytical solutions for the PL transients under various conditions. Specifically, we explore the influence of trap distributions and associated power law transients. Our results indicate that the PL transients can vary as $\Phi_{PL}(t) \propto t^{-\beta}$ for several scenarios with a wide range of values for $\beta$. The results are summarized in Table \ref{tab:power_TPL} and the subsequent sections provide brief theoretical arguments and numerical simulations on such power law transients.\\

\begin{table*}[t]
\centering
\renewcommand{\arraystretch}{2.4}
\setlength{\tabcolsep}{4.5pt}
\small
\captionsetup{labelformat=empty}

\begin{tabular}{|c|c|c|c|c|c|}
\hline
\textbf{Trap type} 
& \textbf{Dominant recombination} 
& $\boldsymbol{p(t)}$ 
& $\boldsymbol{n(t)}$ 
& $\boldsymbol{\Phi_{PL}(t)}$ 
& $\boldsymbol{\beta}$\\
\hline

\multirow{2}{*}{\shortstack{Mid-bandgap\\
recombination centers 
%\\($n = p,\; n_{T} \ll n,p$)
}}
& monomolecular
& $e^{-k_1 t}$
& $e^{-k_1 t}$
& $e^{-2k_1 t}$
& exponential transients\\
\cline{2-6}
& bimolecular
& $t^{-1}$
& $t^{-1}$
& $t^{-2}$
& $\beta$ = 2\\
\hline

\multirow{2}{*}{\shortstack{Uniform %\\ ($n \ll p,\; p \approx n_{T}$)}
}}
& monomolecular
& $\ln(t)$
& $t^{-1}$
& $t^{-1}\ln(t)$ 
& $\beta \approx 1$\\
\cline{2-6}
& bimolecular
& $\ln(t)$
& $t^{-1}$
& $t^{-1}$ 
& $\beta \approx 1$\\
\hline

\multirow{2}{*}{\shortstack{Exponential %\\ ($n \ll p,\; p \approx n_{T}$)
}}
& monomoleular
& $t^{-\alpha k_{B}T}$
& $t^{-(\alpha k_{B}T+1)}$
& $t^{-(2\alpha k_{B}T+1)}$ 
& $\beta = 2k_{B}T/E_U + 1$\\
\cline{2-6}
& bimolecular
& $t^{-\alpha k_{B}T}$
& $t^{-1}$
& $t^{-(\alpha k_{B}T + 1)}$ 
& $\beta = k_{B}T/E_U + 1$\\
\hline

\end{tabular}
\centering
\captionsetup{font=small}
\caption{\justifying Table \ref{tab:power_TPL}.
Trap distributions and corresponding power law exponents.}
\label{tab:power_TPL}

\end{table*}

\textbf{A. Trap free response} The nature of PL transients in the absence of carrier trapping is well known \cite{nair_2023,kirchartz2020photoluminescence}. In the absence of  trapping centers and negligible Auger recombination, the system of equations \ref{eq1}-\ref{eq_p} reduces to $dn/dt=k_1n+k_2n^2$ with $n=p$.  The  general solution is given as $n(t)=k_1/(-k_2+(k_2+k_1n_0^{-1})e^{k_1t})$. The PL signal is given as (see Secs. S2-S3, SI for details)
\begin{equation}
 \Phi_{PL}(t)=\frac{k_2k_1^2n_0^2e^{-2k_1t}}{(k_1+k_2n_0(1-e^{-2k_1t}))^2}
 \label{trap_free}
\end{equation}
where $n_0$ is the carrier concentration at $t=0$. Accordingly, the transients are exponential in nature if $k_1n>>k_2n^2$ with the characteristic decay time being $1/2k_1$. Similarly, the transients show power law trends $t^{-2}$ if $k_2n^2>>k_1n$. Note that $\beta=2$ in this case. A key observation is that power law exponents other than $\beta=2$ are not predicted by this model. In typical experiments, it might take a certain time delay for the power law trends to manifest in the transient PL. During this transition phase, it might be possible to extract $\beta \ne 2$ from experimental results. However, Figure \ref{fig:experiment_combined} shows that $\beta\neq 2$ is clearly evident over several orders of magnitude of time delay in many experimental reports. This indicates that other physical phenomena are also important during PL transients. \\

\textbf{B. Uniform Trap Distribution} We first consider the case of a \textit{uniform trap distribution} (see Fig. ~\ref{uniform_sim}a). In the presence of significant trapping, we have $n_T > n$ and $p\approx n_T$ during the OFF transients. Under such conditions, trapping-detrapping of electrons between the conduction band and trap states has a critical impact on the carrier dynamics. Accordingly, the net rate of electron-hole recombination is given by $\partial p/\partial t$ and not $\partial n/\partial t$ (see eqs. \ref{eq1}-\ref{eq2}). 

The time dependence of $p(t)$ can be arrived at as follows: At any given time $t$, we can assume that traps with emission times $\tau_E < t$ are empty, whereas traps with $\tau_E > t$ remain occupied and continue to influence carrier relaxation dynamics. Here, $\tau_{E} = \tau_{0} \, e^{(E_C-E_{T'})/k_BT} $
where $\tau_{0}^{-1} = c_{n}N_{C}$. Accordingly, the time-dependent trapped carrier density can be expressed as
\begin{equation}
    n_{T}(t) = N_{0}\big(E_{T'}(t) - E_{i}\big)
\end{equation}
where $E_{T'}(t)$ denotes the trap energy level corresponding to an emission time $\tau_{E}=t$ (see Sec. S4, SI, for details). \\

From these relations, it can be shown that the majority carrier (hole) density varies logarithmically with time as
\begin{equation}
    p(t) \propto \ln(t)
\end{equation}
As mentioned before, the net rate of electron-hole recombination is given by $\partial p/\partial t$ and not $\partial n/\partial t$. Hence, the recombination rate varies as $t^{-1}$. However, the recombination could be dominated by either monomolecular or bimolecular processes. This leads to two scenarios, as follows. (i) $k_1n>>k_2np$: Here, we have $n(t)\propto t^{-1}$ and hence $\Phi_{PL}(t) \propto t^{-1}\times ln(t)$. (ii) $k_2np>>k_1n$: Here we have $n(t) \propto t^{-1}/ln(t)$ and hence $\Phi_{PL}(t) \propto t^{-1}$. In both these cases, we find that $\Phi_{PL}(t)\propto t^{-\beta}$ with $\beta \approx 1$. These trends are summarized in Table \ref{tab:power_TPL}.\\

\textbf{C. Exponential Trap Distribution} Figure ~\ref{uniform_sim}b depicts an exponential distribution of traps which can be described as
\begin{equation}
    N_{T}(E) = N_{0} \, e^{-\alpha (E_{C} - E_{T'})}
\end{equation}
where $\alpha^{-1}$ represents the Urbach energy ($E_{U}$), and $(E_{C} - E_{T'})$ denotes the trap depth from the conduction band edge.  At any given time $t$, traps with emission times $\tau_{E} < t$ are expected to be nearly empty, while those with $\tau_{E} > t$ remain significantly occupied. Accordingly, for long term transients, the temporal variation in the hole density is given as (see details in Sec. S6, SI)
\begin{equation}
p(t) \propto t^{-\alpha k_BT}
%    p(t) \propto \frac{N_{0}}{\alpha} \left( \frac{t}{\tau_{0}} \right)^{-\alpha kT}
\end{equation}
As indicated by eqs. \ref{eq1}-\ref{eq2}, the net recombination rate is given as   $R=\partial p/\partial t \propto t^{-(\alpha k_BT+1)}$. Depending on the dominant recombination mechanism, this leads to distinct power law transients. For $k_1n>>k_2n^2$, the above insights lead to $n(t)\propto t^{-(\alpha k_BT+1)}$ and hence $\Phi_{PL}(t)\propto t^{-(2\alpha k_BT+1)}$. On the other hand for $k_2n^2>>k_1n$, we find $\Phi_{PL}(t)\propto t^{-(\alpha k_BT+1)}$. For both cases, we find that $\Phi_{PL}(t) \propto t^{-\beta}$. Depending on the dominant recombination mechanism and the Urbach energy, a broad range of non-integer power law exponents are possible in PL transients of perovskites. 

\section{\label{sec:results} Results and Discussions}

Table \ref{tab:power_TPL} summarizes the trends for both uniform and exponential trap distributions under various conditions. As discused in previous sections, power law transients are expected in the presence of significant trapping-emission. In addition, as indicated by Table \ref{tab:power_TPL}, the power law exponents could be much different from $2$. For example, the model predicts that the power exponent could be $\approx 1$ in the presence of uniform trap distribution - regardless of whether dominated by trap assisted recombination or bimolecular recombination. On the other hand, for exponential distribution of traps, the power exponent could be much larger. Interestingly, for $E_U=2k_BT$, the model predicts that $\beta=2$ even when dominated by monomolecular recombination. This is a surprising result, and is in contrast to the conventional understanding which indicates that $\beta=2$ occurs only when dominated by bimolecular recombination and under conditions $n=p$ ((see eq. \ref{trap_free}).

 \begin{figure*}[t]
    \centering
    \includegraphics[width=\textwidth]{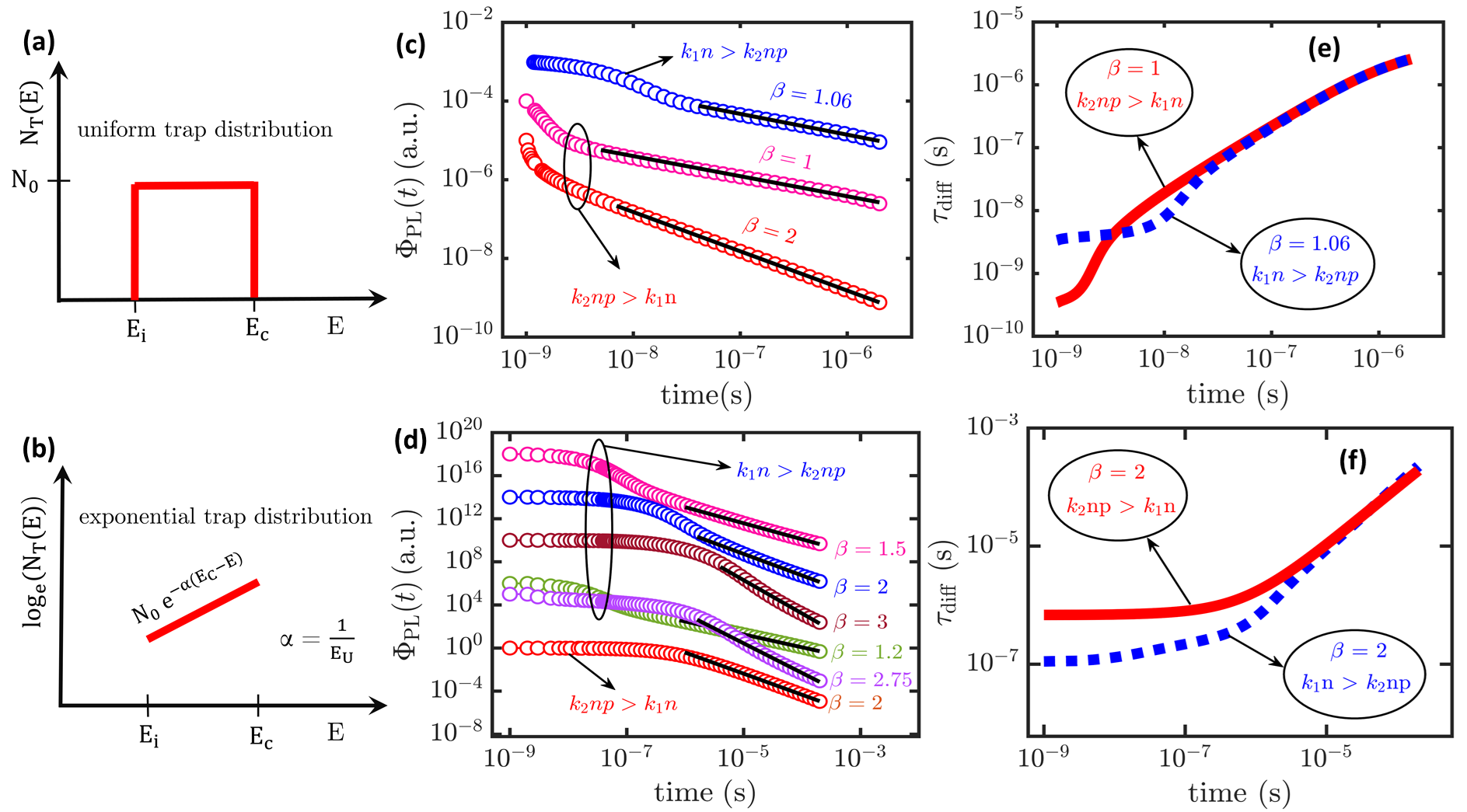} %PCombined_figure_aug7.png
   \captionsetup{labelformat=empty}
    \captionsetup{font=small}
    \caption{\justifying
    \textbf{Figure~\ref{uniform_sim}.}
    Numerical results for different trap distributions and the corresponding transient photoluminescence (TPL) responses.
    (a) Schematic representation of a uniform distribution of traps.
    (b) Schematic representation of an exponential distribution of traps.
    (c) Simulated TPL response for a uniform trap distribution (scaled by a constant factor in y-axis for uncluttered visualization).
    (d) Simulated TPL response for an exponential trap distribution with different Urbach energies (scaled by a constant factor in y-axis for uncluttered visualization).
    (e) Differential decay time, $\tau_{\mathrm{diff}}$, for $\beta \approx 1$ as a function of time for uniform trap distribution.
    (f) $\tau_{\mathrm{diff}}$ versus time (t) for $\beta = 2$ in the presence of exponential trap distribution. For part (d), the extracted power-law exponents are $\beta = 1.2$ for $E_U = 259$ meV, $\beta = 1.5$ for $E_U = 103$ meV, $\beta = 2$ for $E_U = 51.8$ meV, $\beta = 2.75$ for $E_U = 29.4$ meV, and $\beta = 3$ for $E_U = 25.8$ meV, corresponding to the condition $k_1 n > k_2 np$. For $E_U = 25.8$ meV in the bimolecular recombination regime ($k_2 np > k_1 n$), the transient has $\beta = 2$. Additional information on the numerical results are available in Secs. S4-S7, SI. }

    \label{uniform_sim}
\end{figure*}

% ======= Caption appearance =======

Figure \ref{uniform_sim}c and Figure \ref{uniform_sim}d show the numerical simulation results for the PL transients. These results were obtained through simultaneous solution of eqs. \ref{eq1}-\ref{eq_p}. To ensure that the results are not influenced by initial conditions, multiple time cycles of laser pulse excitation and subsequent PL transients are numerically simulated (for further details, see Sections~S5 and~S7 of SI). In accordance with the theoretical expectations, Figure \ref{uniform_sim}c shows that $\beta \approx 1$ for PL transients in the presence of uniform density of traps. Curiously, the model predicts the possibility of $\beta \le 1$, a trend observed in experiments (see Figure \ref{fig:experiment_combined}). Further, such power law trends with $\beta \le 1 $ are not predicted by the traditional analysis of PL transients (see eq. \ref{trap_free}).

PL transients in the presence of an exponential distribution of traps are shown in Figure \ref{uniform_sim}d. As indicated by Table \ref{tab:power_TPL}, $\beta > 1$ in the presence of such Urbach band tail states. In addition, the model predicts that $\beta=2$ under several conditions. For $k_2np>>k_1n$ (i.e., dominant bimolecular recombination), we have $\beta=2$, in accordance with literature. However, interestingly, we find that $\beta=2$ for $E_U=2k_BT$ for $k_1n>>k_2np$. This indicates that observation of power law transients with $\beta=2$ does not imply that the bimolecular process is the dominant recombination. On the other hand, in the presence of significant trapping/emission, transients with $\beta=2$ are possible even when monomolecular recombination is dominant (i.e., $k_1n>k_2np$). Additional details on this interesting result are provided in  Fig. S7.1 and Fig. S7.2 of SI. As predicted by the model (see also Table \ref{tab:power_TPL}), our numerical simulations indicate that $\beta>2$ for $E_U<2k_BT$ and $\beta <2 $ for $E_U> 2k_BT$ with monomolecular recombination as the dominant mechanism. Similar insights regarding $\beta$ and $E_U$ are readily available from  Table \ref{tab:power_TPL} for other scenarios. Details regarding the temporal variations in $n$, $p$, $n_T$ and the recombination/trapping/emission rates for each of the above discussed scenarios are provided in SI.

Our results indicate a clear correlation between trap distributions and PL transients. For example, samples with a lower Urbach energy, indicative of reduced energetic disorder, exhibit a faster PL decay. In contrast, samples with higher $E_{U}$ values display slower decay characteristics due to the increased contribution of deeper or more broadly distributed trap states. These findings highlight the intricate interplay between energetic disorder, trap distribution, and recombination kinetics that shapes the nonexponential power law transients in halide perovskites.

With the possibility and relevance of power law transients conclusively established through theoretical methods, numerical simulations, and experimental observations, we now focus on the implications of the same on derived concepts like differential carrier lifetime. Literature indicates that the relative dominance of recombination mechanisms can be understood in terms of the differential carrier lifetime given as \cite{kirchartz2020photoluminescence,yuan2025understanding}
\begin{equation}
    \tau_{\mathrm{diff}}=\left(-\frac{1}{m}\frac{d\ln \Phi_{\mathrm{TPL}}}{dt}\right)^{-1}
    \label{eq:difft}
\end{equation}
For example, in the absence of any trapping/emission, with $n=p$, and $k_1n>>k_2n^2$, we have $\Phi_{PL}(t) \propto e^{-2k_1t}$. Eq. \ref{eq:difft} with $m=2$ indicates that $    \tau_{\mathrm{diff}} =\mathrm{constant}$ in this case. %Eq. \ref{eq:difft} with $m=2$ indicates that $    \tau_{\mathrm{diff}}=k_1^{-1}$ - the electron recombination time. 
With dominant bimolecular recombination, we have $\Phi_{PL}(t) \propto t^{-2}$. Here, we find that $\tau_{\mathrm{diff}}$ increases linearly with time. These trends are often used to identify the dominant recombination mechanism -i.e., a linear increase in $\tau_{\mathrm{diff}}$ indicates that the dominant recombination mechanism is bimolecular.

The above well-established insight, curiously, is invalid in the presence of power-law transients. With $\Phi_{PL}(t) \propto t^{-\beta}$, eq. \ref{eq:difft} indicates that $\tau_{\mathrm{diff}} \propto t$ - i.e., $\tau_{\mathrm{diff}}$ is expected to vary linearly with $t$ for any $\beta$. Our analysis and the results shared in Figs. \ref{uniform_sim}c,d indicate that power-law transients are possible regardless of the dominant recombination mechanism. Hence, an observation of a linear increase in $\tau_{\mathrm{diff}}$ is not a rigorous metric to identify the dominant recombination mechanism. This argument is supported through numerical simulation results shown in Figs. \ref{uniform_sim}e,f. Here, the $\tau_{\mathrm{diff}}$ for different scenarios are plotted - part (e) for uniform trap distribution and part (f) for exponential trap distribution. In both cases, we find that $\tau_{\mathrm{diff}}$ varies linearly with time - regardless of the dominant recombination mechanism.

\section{Conclusions}

Here, we theoretically establish that the PL transients could vary as $t^{-\beta}$ in perovskites (with $\beta \neq 2$) due to carrier trapping/emission. Our models indicate that the power law exponent $\beta$ is heavily influenced by the nature of the trap distribution and explain the wide range of $\beta$ reported in the literature. We also elucidate the shortcomings associated with analysis of PL transients in terms of multi-exponentials with the data plotted in linear $y$ vs. linear $x$ format. As such, this work  highlights the need to combine physics based predictive modeling with insightful data analysis to unravel the working mechanisms of perovskite based optoelectronic devices.

%\vspace{0.5cm}

% ================================
% Acknowledgements and Statements
% ================================
\textbf{Acknowledgements}: The authors acknowledge the National Center for Photovoltaics Research and Education (NCPRE), 
Indian Institute of Technology Bombay, for research support.

\section*{Supplementary Information}

Additional discussions on the analytical derivation of carrier dynamics for both uniform and exponential trap distributions are provided in the Supplementary Information. 
Further, additional details regarding simulation results which includes recombination dynamics, carrier dynamics, transient photoluminescence (TPL) behavior, differential decay, and quasi-Fermi levels are also provided in the Supplementary Information. \\

\section*{Conflict of Interest}
The authors declare that they have no conflict of interest.

\section*{Data Availability}
The data that supports the findings of this study are available within the article and in the Supplementary Information.\\

\section{\label{sec:two} References}
% =====================================================
% References (in two-column continuation)
% =====================================================
%\bibliographystyle{IEEEtran}  % IEEE two-column reference style
\bibliography{references}     % Your BibTeX file (references.bib)

\end{document}